\documentclass[journal,10pt,twocolumn]{IEEEtran}
\usepackage{multirow}
\usepackage[algoruled, linesnumbered]{algorithm2e}
\usepackage{cite}
\usepackage{graphicx}
\usepackage{psfrag}
\usepackage{epsfig}
\usepackage{subfigure}
\usepackage[hyphens]{url}
\usepackage{amsmath}
\usepackage{array}
\usepackage{wasysym}
\usepackage{hyperref}

\begin{document}

\title{Comparison of Speech Activity Detection Techniques for Speaker Recognition}

\author{Md~Sahidullah,~\IEEEmembership{Student Member,~IEEE,}
        Goutam~Saha,~\IEEEmembership{Member,~IEEE}

\thanks{The authors are with the Department of Electronics and Electrical Communication Engineering, Indian Institute of Technology Kharagpur, Kharagpur, West Bengal, 721302, India. e-mail: sahidullahmd@gmail.com; gsaha@ece.iitkgp.ernet.in}}

\maketitle

\begin{abstract}
Speech activity detection (SAD) is an essential component for a variety of speech processing applications. It has been observed that performances of various speech based tasks are very much dependent on the efficiency of the SAD. In this paper, we have systematically reviewed some popular SAD techniques and their applications in speaker recognition. Speaker verification system using different SAD technique are experimentally evaluated on NIST speech corpora using Gaussian mixture model- universal background model (GMM-UBM) based classifier for clean and noisy conditions. It has been found that two Gaussian modeling based SAD is comparatively better than other SAD techniques for different types of noises.
\end{abstract}


\section{Introduction}\label{Section:Introduction}
Speech activity detection (SAD) is an important task in most of the speech processing applications. The function of a SAD is to distinguish silence, non-speech frames from speech signals. It has been found that the presence of non-speech frames considerably affects performance of the system. Speech activity detection is also called voice activity detection. However, as in speech processing terminology voice and speech are not same, we call the task of identifying speech frames as SAD throughout this work. SAD techniques are designed using various methods. Most of them use heuristically chosen statistical properties of speech parameters like: energy, pitch, entropy etc. Therefore, the performance of different SAD are different and varying according to the level and type of signal-to-noise ratio (SNR). As a result, the performances of different speech based systems are significantly sensitive to the employed SAD technique. Therefore, SAD should be carefully chosen while designing a speech based system. Speech activity detection is rigorously studied for speech recognition, speech coding etc. However, it is not so far thoroughly studied for speaker recognition applications. Very recently, it has drawn attention of the researchers in this field~\cite{PadmanabhanThesis,ParthasarathiPadmanabhan,ZaurNasibov}.
\par
In current speaker recognition systems, energy based SADs are predominantly used. For example, Kinnunen \textit{et. al} has employed energy based SAD which is found useful for NIST speech data~\cite{Kinnunen}. The baseline speaker recognition system developed in~\cite{PrasannaPradhan} also uses a different energy based SAD. A bi-Gaussian model of speech frame's log energy distribution is suggested in~\cite{bimbot1}. In recent days, different SADs are used for different quality of speech signal. For example, in~\cite{DehakKennyIvector}, for speech frame selection Hungarian phoneme recognizer developed at BUT is used for telephone quality speech, and on the other hand, GMM based approach is used  for microphone quality speech~\cite{Hanwu}. Other than this, voice activity detector used in \textsc{G.729B}, statistical voice activity detector proposed by Sohn \textit{et. al} are also used for speech activity detection in speaker recognition. In this paper, we briefly review all those techniques. Then, their performances are compared for two popular NIST speech corpus for clean and noisy environmental conditions. The performances is also evaluated on a simulated real-time situation where speech utterances of training and testing are distorted with various noises of different SNRs. In most of the cases, we have found that speaker recognition systems with two Gaussian modeling based SADs are significantly better than other techniques.
\par
The rest of the paper is organized as follows. In Section~\ref{Section:Review}, the existing speech activity detectors are briefly reviewed. The experimental setup used in this paper is discussed in Section~\ref{Section:ExperimentalSetup}. The results obtained in different experiments are discussed in Section~\ref{Section:ResultsDiscussion}. Finally, the paper is concluded in Section~\ref{Section:Conclusion}.

\section{Review of Some Popular SAD Techniques}\label{Section:Review}
Most of the speech activity detectors are based on either time domain or frequency domain approach. Various time domain features like \textit{short-time average energy (STAE)}, \textit{short-time average magnitude (STAM)}, \textit{zero-crossing rate (ZCR)} and so on are used in time domain. On the other hand, in frequency domain, various spectral information are used for designing a SAD. There are numerous examples where these time and frequency domain information and their statistical properties are used to develop robust speech activity detector~\cite{Benyassine,Sohn,SADTanyerOzer,SADRamirezSegura,SADChoKondoz,SADChangKimMitra}. Speech activity detectors based on periodicity measure of speech signal is used in~\cite{SADTucker}. Cepstral information based SAD is proposed in~\cite{HaighMason}. In~\cite{SADrez}, SAD based on long term speech information is proposed for automatic speech recognition. Transformed domain characteristics of speech signal are used to design SAD in~\cite{SADGazorZhang}. Entropy based SAD is proposed in~\cite{SADEntropyRenevey}. Divergence of subband information is utilized in~\cite{SADRamirezKLD}. Recently, modulation spectrum information in terms of delta-phase spectrum is used to design robust voice activity detection for robust speaker recognition~\cite{SADDeltaPhase}.
\par
A concise and updated review of the existing SAD techniques are not available yet. However, some older survey exists in this domain. For example, a class of frequency domain voice activity detector used for VoIP speech compression is compared in~\cite{CompSADVoIP}. In~\cite{CompSADThree}, three popular voice activity detectors in speech coding domain are experimentally evaluated and assessed using different subjective and objective indices. VADs are compared in wireless application domain in~\cite{CompSADWireless}. As there are enormous amount of work in this domain it is quite difficult to evaluate all of them in a single work. In this paper, we review five popular SAD techniques which are commonly used in state-of-the art speaker recognition system.

\subsection{SAD Used in G.729B~\cite{Benyassine}}\label{Subsection:G729B}
Voice activity detection used in G.729 codec is widely used in different speech processing applications. This technique classifies active and inactive voice frames with the help of multiple speech parameters in the following steps:

\begin{itemize}
  \item
  First, four parameters: line spectral pairs (LSFs), full-band energy, low-band energy and the zero-crossing rate are computed for different speech frames of the given speech signal.
  \item
  Parameters of initial frames of the signals are considered as background noise information and distortion is measured with respect to those background parameters.
  \item
  Initial VAD decisions are made using \emph{multiboundary} of four parameter set.
  \item
  Initial VAD decisions are smoothed in four steps in order to reflect stationary nature of speech and background signal.
\end{itemize}

The VAD algorithm is terminated after the estimated background noise information crosses a pre-defined threshold.

\subsection{Statistical SAD~\cite{Sohn}}\label{Subsection:SSAD}
In this SAD technique, a statistical model is used where the decision are taken based on likelihood ratio test (LRT). In addition to that, a decision-directed (DD) method is used to estimate various parameters. Finally, it also introduces an improved hang-over scheme based on hidden Markov model (HMM). The purpose of hang-over scheme is to detect the speech frames which almost buried in noise. It has been shown that this approach performs significantly better than G.729B based SAD in low SNR for speech frame detection.

\subsection{Energy Based SAD~\cite{Kinnunen,PrasannaPradhan}}\label{Subsection:ESAD}
Energy based SAD techniques are very straightforward, and they are widely used in speech and speaker recognition application. First, energy of all the speech frames are computed for a given speech utterance. Then, an empirical threshold is selected from the frame energies. In~\cite{Kinnunen}, the threshold is determined from the maximum energy of the speech frames. In~\cite{PrasannaPradhan}, the threshold is selected as $0.06 \times E_{avg}$, where $E_{avg}$ is the average energy of the frames a speech utterance. These kinds of techniques are somewhat suitable for clean condition. But, the performance of the system degrades significantly in low SNR.

\subsection{Phoneme Recognizer Based SAD~\cite{Schwarz}}\label{Subsection:PSAD}
State-of-the art speaker recognition system uses \textit{Hungarian phoneme recognizer tool}\footnote[1]{\url{http://speech.fit.vutbr.cz/software/phoneme-recognizer-based-long-temporal-context}} to mark speech and non-speech frames for telephone quality speech signals~\cite{GlembekBurget,Alam2012}. This tool identifies speech segments as speech or non-speech segments. The speech segments are assigned to different phonemes. On the other hand, non speech segments are of four kinds\footnote[2]{\url{http://www.fee.vutbr.cz/SPEECHDAT-E/public/Deliver/WP1/ED141v11-fin.doc}}: (i)\textit{pau}: pause within speech signal, (ii)\textit{spk}: speaker related noises, (iii)\textit{sta}: stationary noise, and (iv)\textit{int}: intermittent noise. This tool uses a neural network and hidden Markov model based technique to identify phonemes. Neural network is used to train speech frames according to their target labels which is usually obtained from a standard database.

\subsection{Bi Gaussian Modeling of log-Energies Based SAD~\cite{bimbot1}}\label{Subsection:BiGaussian}
\begin{figure}
  \centering
  \includegraphics[width=9cm]{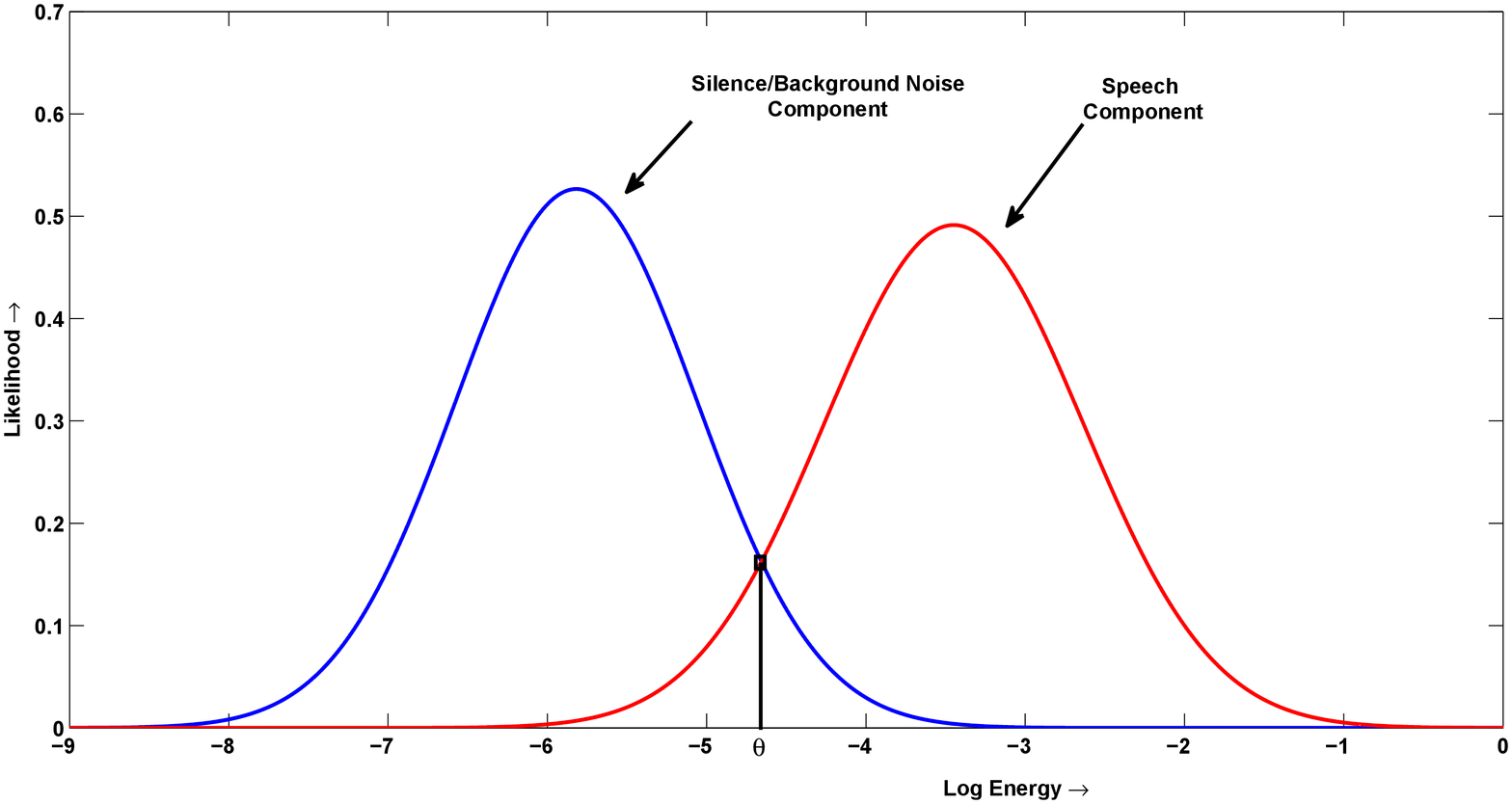}\\
  \caption{Bi Gaussian model of log-energy values of speech frames. Gaussian distribution function with 'Red' color represents speech component and 'Blue' color represents noise/background component. Here, $\theta$ is the threshold.}\label{FigBiGaussian}
\end{figure}

Bi-Gaussian modeling for speech frame selection is briefly explained in~\cite{bimbot1,MagrinSAD}. In this method, first the log-energies of each speech frames of a speech utterance are computed. Then the distribution of log-energy coefficients is estimated using Gaussian mixture model of two mixtures. The cluster corresponding to smaller value of center is treated as noise or non-speech class, and the cluster corresponding to larger value of center is considered as speech (Fig.~\ref{FigBiGaussian}). A threshold is computed to determine the decision making boundary between speech and non-speech class. Usually, it is chosen as the point between the two centers where the probabilities are equal.

\section{Experimental Setup}\label{Section:ExperimentalSetup}

\subsection{Database Description}\label{Subsection:Database}
Speaker recognition experiments have been conducted on NIST SRE $2001$~\cite{MarkPrzybocki} and NIST SRE $2002$~\cite{SreEvaluationPlan:2002}. In literature, it has been found that noise related experiments are mostly performed on those database. These two datasets have less variability due to channel and handset compared to the latest NIST corpora. Therefore, in order to observe the effect in presence of adverse environmental condition, synthetic addition of noise to this dataset is more justified. We have performed all the experiments on core task condition of the evaluation plan. The detail description of the database is shown in Table~\ref{TableDatabase}. In both the cases, we have used the development data of NIST SRE $2001$ for UBM preparation.

\begin{table}
\renewcommand{\arraystretch}{1.3}
\caption{Database description (coretest section) for the performance evaluation of various speech activity detectors.}
\label{TableDatabase}
\centering
\begin{tabular}{|c|c|c|}
\hline
 & SRE 2001  & SRE 2002               \\
\hline     \hline
Target Models      &  74\male, 100\female &  139\male, 191\female  \\
\hline
Test Segments                  &  2038       &  3570\\
\hline
Total Trial                    &  22418      &  39270\\
\hline
True Trial                     &  2038       &  2983\\
\hline
Impostor Trial                 &  20380      &  36287\\
\hline
\end{tabular}
\end{table}

\subsection{Feature Extraction}\label{Subsection:FeatureExtraction}
MFCC features have been extracted using $20$ filters linearly spaced in mel scale from speech frame of $20$ms keeping $50\%$ overlap with adjacent frames. The details of our feature extraction process can be found in~\cite{SahidullahBlockTransform}. We have also shown the performance for our previously proposed overlapped block transform coefficient (OBTC) feature termed as OBT-9-13 which extracted by performing distributed DCT on the mel filterbank output~\cite{SahidullahBlockTransform}. The dimensions of the MFCC and OBTC features are $38$ and $40$ correspondingly after static features are augmented with their velocity coeffificents.

\subsection{Classifier Description}\label{Subsection:Classifier Description}
In this paper, all the performance are based on GMM-UBM based classifier where the target models are created by adapting UBM parameters~\cite{Reynoldsadapted}. Initially, a gender dependent UBM with $256$ mixtures is trained using expectation-maximization algorithm after initialization using split vector quantization with data from development section of NIST SRE $2001$. Target models are created by adapting the centres of the UBM with relevance factor of $14$. Finally, during score computation, only top-$5$ Gaussians of UBM per frame are considered.

\subsection{Performance Evaluation Metric}\label{Subsection:Metric}
The performance of speaker recognition is measured using equal error rate (EER) metric which is a particular operating point on detection error tradeoff plot where the probability of false rejection (FR) equals probability of false acceptance (FA). The performance is also measured in terms of minimum detection cost function (minDCF) where a cost function is minimized by assigning unequal cost to FR and FA. Here, the costs of FR and FA are set at $1$ and $10$ correspondingly.

\section{Results and Discussions}\label{Section:ResultsDiscussion}
Speaker recognition experiments are conducted on both the databases using MFCC and OBTC feature. In all the experiments, everything other than the SAD techniques are kept fixed to observe the effect of SAD. In this paper, the SAD techniques are abbreviated as follows:
\begin{itemize}
  \item
  \textbf{G.729B}: SAD used in G.729B~\cite{Benyassine}.

  \item
  \textbf{SMSAD}: Statistical model based SAD proposed by Sohn \textsc{et. al}~\cite{Sohn}.

  \item
  \textbf{HNPNR}: Hungarian phoneme recognizer based SAD~\cite{Schwarz}.

  \item
  \textbf{MEBTS}: Energy dependent SAD using maximum energy based threshold selection~\cite{Kinnunen}.

  \item
  \textbf{AEBTS}: Energy dependent SAD using average energy based threshold selection~\cite{PrasannaPradhan}.

  \item
  \textbf{UBGME}: Utterance-wise bi Gaussian modeling of log-energy based threshold selection~\cite{MagrinSAD}.
\end{itemize}

\subsection{Results in Match Condition}\label{Subsection:MatchCondition}
Experiments are first conducted on clean speech database. Speaker verification results with different SADs are shown in Table~\ref{TableCleanCompare}. Frame selection technique which is used in \textsc{G.729B} is shown to perform worst for speaker recognition in almost all the cases. The performance of SMSAD is better than the performance of \textsc{G.729B} for most of the cases. HNPNR based SAD is shown to outperform the first two techniques. However, MEBTS based approach is consistently better than those techniques for all the cases. Then, bi Gaussian modeling based approach (UBGME) is shown to perform better than all the previously mentioned techniques for NIST SRE $2001$. However, its performance is slightly reduced than HNPNR, MEBTS and AEBTS for NIST SRE $2002$. It can also be observed that energy based SADs are relatively better than the first two techniques. Though these two techniques are very useful for speech coding and compression, but they are not much important in speaker recognition. This is most likely due to the fact that those techniques consider a signal frame as speech if it has some speech-like information. However, the energy based SADs consider a frame as speech frame if only if it has significant amount of energy. Though low energy speech frames may have speech information, but their contribution in speaker recognition seems to be negligible.
\par
The overall energy of the frames increases when the speech is distorted by noise in adverse conditions. In that case, merely energy of the frame seems to be unreliable for speech activity detection. Therefore, it is worth studying the effect of the SAD in speaker recognition for noisy condition. This study is carried out on both the databases and the same is discussed in the following subsection.

\begin{table*}
\renewcommand{\arraystretch}{1.3}
  \centering
  \caption{Speaker verification results in terms of equal error rate and minimum detection cost function on NIST SRE 2001 and NIST SRE 2002 for different implementations of speech activity detector. [\textbf{G.729B}: SAD used in G.729 Codec~\cite{Benyassine}, \textbf{SMSAD}: Statistical Model based SAD~\cite{Sohn}, \textbf{HNPNR}: Hungarian Phoneme Recognizer~\cite{Schwarz}, \textbf{MEBTS}: Maximum Energy Based Threshold Selection~\cite{Kinnunen}, \textbf{AEBTS}: Average Energy Based Threshold Selection~\cite{PrasannaPradhan}, \textbf{UBGME}: Utterance Wise Bi Gaussian Modeling of Log Energy~\cite{MagrinSAD}.]}
  \begin{tabular}{|c||c|c|c|c||c|c|c|c|}
     \hline
     \hline
     \multirow{3}{*}{Method}   & \multicolumn{4}{|c|}{NIST SRE 2001} & \multicolumn{4}{|c|}{NIST SRE 2002}\\
     \cline{2-9}
                               & \multicolumn{2}{|c|}{EER (in \%)} & \multicolumn{2}{|c|}{minDCF $\times$ 100} & \multicolumn{2}{|c|}{EER (in \%)} & \multicolumn{2}{|c|}{minDCF $\times$ 100}\\
     \cline{2-9}
                               &  MFCC  & OBTC  &  MFCC  & OBTC&  MFCC  & OBTC  &  MFCC  & OBTC\\
     \hline
     \hline
     G.729B                    &  9.32 & 8.59 & 4.11 & 3.85 & 10.16 & 9.83 & 4.50 & 4.38\\
     \hline
     SMSAD                     &  8.97 & 8.11 & 3.92 & 3.74 &  9.86 & 9.85 & 4.51 & 4.29\\
     \hline
     HNPNR                     &  8.54 & 7.76 & 3.74 & 3.48 &  9.09 & 8.78 & 4.43 & 4.35\\
     \hline
     MEBTS                     &  8.24 & 7.27 & 3.58 & 3.44 &  9.09 & 8.58 & 4.45 & 4.10\\
     \hline
     AEBTS                     &  7.96 & 7.21 & 3.61 & 3.49 &  9.45 & 8.59 & 4.44 & 4.22\\
     \hline
     UBGME                     &  7.91 & 7.36 & 3.72 & 3.41 &  9.69 & 9.05 & 4.50 & 4.28\\
     \hline
     \hline
  \end{tabular}
  \label{TableCleanCompare}
\end{table*}

\subsection{Results in Mismatch Condition}\label{Subsection:MismatchCondition}
The experiments on noisy conditions are performed by synthetically adding noise to the test utterances. The noise samples are taken from NOISEX-92 database\footnote[4]{http://www.speech.cs.cmu.edu/comp.speech/Section1/Data/noisex.html} and they are down-sampled to $8$kHz. The noise is added to the speech signal using the following steps:

\begin{itemize}
  \item
  A segment of noise sample from the original noise signal is randomly selected according to the length of the speech signal with whom noise is to be added.

  \item
  The amplitude of the noise segment is scaled depending on desired SNR.

  \item
  The scaled noise signal is added to the clean signal to get the distorted speech.
\end{itemize}

\begin{figure*}
  \centering
  \includegraphics[width=18cm]{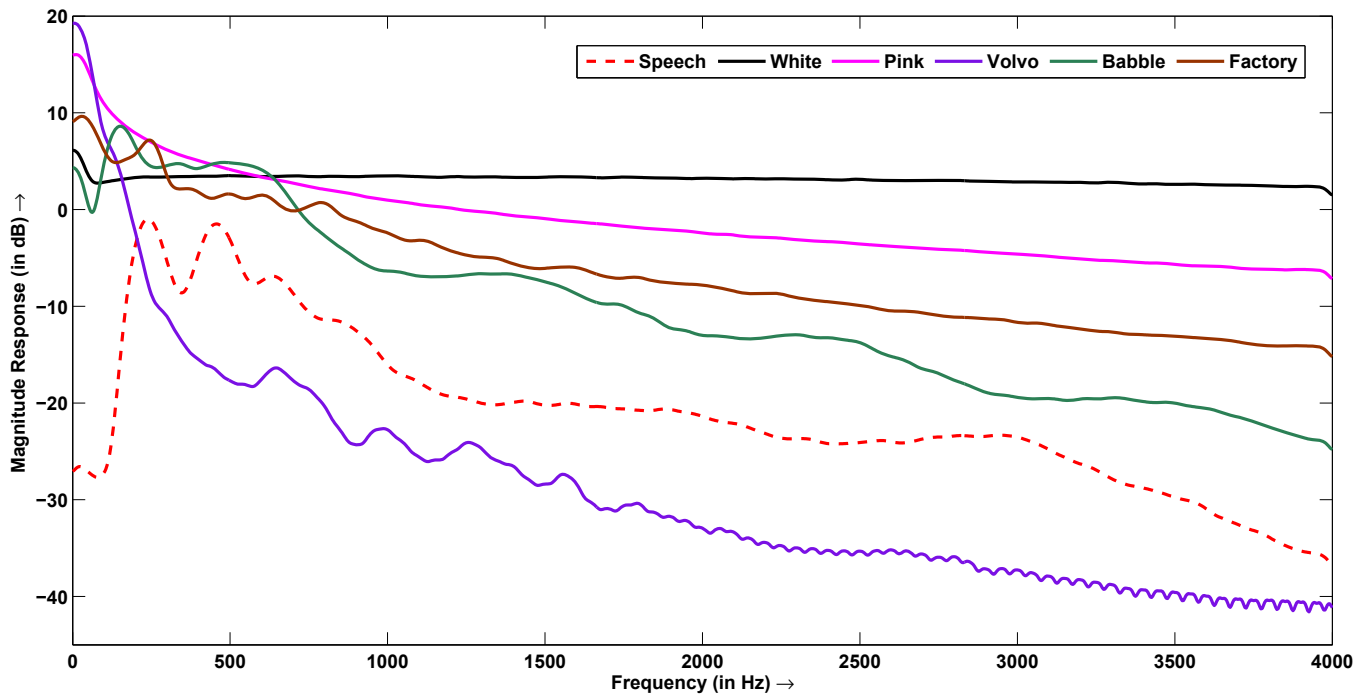}\\
  \caption{Average of short term magnitude responses for clean speech and noise samples. The speech signals are taken from NIST SRE 2001 where noise samples are taken from NOISEX-92 database. Averaging operation is performed over all available signal frames.}\label{FigFrequencyResponses}
\end{figure*}

In our experiment, we have chosen five different noise samples: \textit{white}, \textit{pink}, \textit{volvo}, \textit{babble} and \textit{factory} (Factory-1 noise in the original database). We have chosen those noises due to their various frequency domain behavior as shown in Fig.~\ref{FigFrequencyResponses}. The experiments are conducted for three different levels of noise: \textit{high} ($0$ dB), \textit{medium} ($10$ dB) and \textit{low} ($20$ dB). The results are shown in Table~\ref{TableNoiseSRE2001} and Table~\ref{TableNoiseSRE2002} for the two databases.

\par
Performance of the speaker recognition system drops severely in presence of noise for different types of SAD. The degradation in performance is dependent on the frequency response of the noise. For example, from Fig.~\ref{FigFrequencyResponses}, we can interpret that as the frequency response of the \textsc{white} noise is flat, all the frequency component of the speech signal is affected. Therefore, the performance is worst for this noise. On the other hand, volvo noise affects only first few Mel filter-bank output. Hence, performance in the presence of \textsc{volvo} noise is relatively less degraded compared to other noises of same SNR.
\par
We also note that performance is varying significantly for different SAD techniques in presence of noise. In case of G.729B and SMSAD, the performances are nearly equivalent for both the databases. HNPNR based voice activity detector which is used in state-of-the art speaker recognition system for speech frame selection performs poorly compared to the other techniques. The performance of energy thresholded SADs i.e. MEBTS and AEBTS suffer severely in presence of noise. The performance is even worse compared to G.729B and SMSAD. In presence of noise, all the speech frames are affected i.e. energy of each frames are increased and the frequency response of all the speech frames are severely affected. In this scenario, maximum energy or average energy based threshold selection techniques will not be much effective. Hence, vowel like regions are only seems to be relevant for those cases~\cite{PrasannaPradhan}. The two Gaussian model based approach selects the energy threshold as the boundary between speech and non-speech class which selects vowel like higher energy frames for most of the part. Therefore, the performance is significantly better for this technique in higher noise.

\begin{table*}
\renewcommand{\arraystretch}{1.3}
  \centering
  \caption{Speaker verification results on NIST SRE 2001 in presence of various additive environmental noise. The results are shown for various speech activity detection techniques.}
  \begin{tabular}{|c|c|c|c|c|c|c|c||c|c|c|c|c|c|}
     \hline
     \hline
     \multirow{3}{*}{Method} & \multirow{3}{*}{Noise} & \multicolumn{6}{|c}{EER (in \%)} & \multicolumn{6}{|c|}{minDCF $\times$ 100}\\
     \cline{3-14}
     & & \multicolumn{3}{|c|}{MFCC} & \multicolumn{3}{|c|}{OBTC}& \multicolumn{3}{|c|}{MFCC} & \multicolumn{3}{|c|}{OBTC}\\
     \cline{3-14}
     &&  20 dB & 10 dB  & 0 dB &  20 dB & 10 dB  & 0 dB&  20 dB & 10 dB  & 0 dB &  20 dB & 10 dB  & 0 dB\\
     \hline
     \hline
     \multirow{4}{*}{G.729B} & White   & 11.38  & 16.69 & 28.56 & 10.40 & 15.65 & 27.28 & 4.80  & 7.17 & 9.84 & 4.52  & 7.12 & 9.84\\
     \cline{2-14}
     &Pink     & 10.70  & 13.64 & 23.46 & 9.52 & 12.90 & 22.77 & 4.59  & 6.45 & 9.37 & 4.32 & 6.05 & 9.04\\
     \cline{2-14}
     &Volvo    & 11.30  & 11.43 & 13.14 & 10.60 & 10.50 & 12.07 & 4.96  & 5.03 & 6.21 & 4.73& 4.71 &  5.22\\
     \cline{2-14}
     &Babble   & 9.96  &  13.74 & 23.99 &  9.43 & 12.70 & 22.93 & 4.63   & 6.52 & 9.17 & 4.13  & 5.75 & 8.84\\
     \cline{2-14}
     &Factory  & 10.55  & 13.35 & 20.96 &  9.57 & 12.41 & 19.47 & 4.66  & 6.33 & 8.96 & 4.36 & 5.84 & 8.58\\
     \hline
     \hline
     \multirow{4}{*}{SMVAD} & White   & 11.24  & 15.75 & 27.09 & 9.91 & 14.78 & 27.49 & 4.79  & 7.00 & 9.38 & 4.39  & 6.98 & 9.61\\
     \cline{2-14}
     &Pink     & 10.12  & 13.35 & 23.60  &  9.18 & 12.32 & 22.62 & 4.52  & 6.34 & 9.17 & 4.18  & 5.92 & 9.02\\
     \cline{2-14}
     &Volvo    & 10.85  & 11.13 & 13.20  & 10.16 & 10.26 & 11.97 & 4.79  & 5.05 & 5.89  & 4.62 & 4.55 & 5.11\\
     \cline{2-14}
     &Babble   & 10.16  & 13.82 & 24.92  &  8.92 & 11.83 & 22.87 & 4.62  &  6.44 & 9.22 & 4.08 & 5.55 & 8.86\\
     \cline{2-14}
     &Factory  & 10.06  & 13.05 & 21.19  &  9.27 & 11.63 & 19.28 & 4.55  & 6.03 & 8.70 & 4.28  & 5.67 & 8.42\\
     \hline
     \hline
     \multirow{4}{*}{HNPNR} & White   & 11.19 & 16.44 & 34.64 & 9.81  & 15.02 & 34.49 & 4.99  & 7.41 & 10.00 & 4.58 & 7.35 & 10.00\\
     \cline{2-14}
     &Pink     & 9.87 & 13.15 & 29.94 & 8.83 & 12.41 & 28.70& 4.63  & 6.46  & 10.00  & 4.30 & 6.21  & 10.00\\
     \cline{2-14}
     &Volvo    & 9.67 & 9.86 & 12.32 &8.93 & 9.33 & 10.84 & 4.62  & 4.65 & 5.80 & 4.16  & 4.21 & 4.98\\
     \cline{2-14}
     &Babble   & 9.43 & 12.95 & 24.14 & 8.54 & 11.68 & 23.70 & 4.29  & 6.06 & 9.69 & 3.92  &5.64 & 9.92\\
     \cline{2-14}
     &Factory  & 9.62 & 12.37 & 26.36 & 9.18 & 11.89 & 26.10 & 4.54  & 6.26 & 10.00 & 4.25  & 5.88 & 10.00\\
     \hline
     \hline
     \multirow{4}{*}{MEBTS} & White   & 13.64  & 20.85 & 33.57 & 12.31 & 20.36 & 34.20 & 5.98  & 9.02 & 10.00 & 5.50 & 8.63 & 9.98\\
     \cline{2-14}
     &Pink     & 10.06  & 17.08 & 30.86 & 8.64 & 17.17  & 30.47 & 4.38  & 8.31 & 10.00 & 3.93 & 7.94 & 9.95\\
     \cline{2-14}
     &Volvo    & 8.73  & 9.13 & 10.74 & 7.91 & 8.34 & 9.29 & 3.99   & 4.05 & 4.81 & 3.75 & 3.71 & 4.16\\
     \cline{2-14}
     &Babble   & 8.64  & 13.63 & 29.63 & 7.60 & 13.20 & 29.64 & 3.87  & 6.65 & 9.88 & 3.47 & 6.18 & 9.81\\
     \cline{2-14}
     &Factory  & 9.22  & 15.55 & 29.92 & 8.54 & 15.75 & 28.56 & 4.23 &  7.51 & 10.00 & 3.94 & 7.19 & 9.93\\
     \hline
     \hline
     \multirow{4}{*}{AEBTS} & White   & 11.97  & 21.41 & 34.15 & 10.40 & 19.68 & 33.32 & 5.12  & 8.96 & 10.00  & 4.52 & 8.05 & 10.00\\
     \cline{2-14}
     &Pink     & 9.09  & 17.47 & 31.45 & 7.90 & 16.29 & 29.78 & 4.20  & 8.06 & 10.00 & 3.79 & 7.15 & 9.98\\
     \cline{2-14}
     &Volvo    & 8.58  & 8.99  & 10.30 & 7.89 &  8.25 &  9.18 & 4.00  & 3.95 & 4.62 & 3.69 & 3.66  & 4.10\\
     \cline{2-14}
     &Babble   & 8.29  & 12.22 & 29.70 & 7.57 & 11.38 & 29.34 & 3.79  & 5.90 & 9.93 &  3.46 & 5.51 & 9.84\\
     \cline{2-14}
     &Factory  & 8.88  & 14.62 & 30.57 & 8.10 & 13.94 & 28.60 & 4.08  & 6.80 & 10.00 & 3.76 & 6.26 & 9.98\\
     \hline
     \hline
     \multirow{4}{*}{UBGME} & White & 9.76  & 14.18 & 24.88 & 9.31 & 13.98 & 24.88 & 4.66  & 6.70 & 9.29 & 4.19 & 6.52 & 9.47\\
     \cline{2-14}
     &Pink     & 8.89  & 12.03 & 21.44 & 8.69 & 11.29 & 20.80 & 4.17  & 5.71 & 8.79 & 3.80 & 5.41 & 8.85\\
     \cline{2-14}
     &Volvo    & 8.73  & 9.37  & 11.19 & 8.30 &  8.82 &  9.96 & 4.16  & 4.09  & 4.78  & 3.89  & 3.80  & 4.26\\
     \cline{2-14}
     &Babble   & 8.78  & 11.63 & 20.70 & 8.34 & 10.50 & 20.61 & 3.85  & 5.39 & 8.58 & 3.52 & 5.10 & 8.59\\
     \cline{2-14}
     &Factory  & 9.03  & 11.39 & 21.20 & 8.59 & 11.14 & 20.50 & 4.05  & 5.53 & 8.62 & 3.76  & 5.11 & 8.60\\
     \hline
     \hline
  \end{tabular}
  \label{TableNoiseSRE2001}
\end{table*}

\begin{table*}
\renewcommand{\arraystretch}{1.3}
  \centering
  \caption{Speaker verification results on NIST SRE 2002 in presence of various additive environmental noise. The results are shown for various speech activity detection techniques.}
  \begin{tabular}{|c|c|c|c|c|c|c|c||c|c|c|c|c|c|}
     \hline
     \hline
     \multirow{3}{*}{Method} & \multirow{3}{*}{Noise} & \multicolumn{6}{|c}{EER (in \%)} & \multicolumn{6}{|c|}{minDCF $\times$ 100}\\
     \cline{3-14}
     & & \multicolumn{3}{|c|}{MFCC} & \multicolumn{3}{|c|}{OBTC}& \multicolumn{3}{|c|}{MFCC} & \multicolumn{3}{|c|}{OBTC}\\
     \cline{3-14}
     &&  20 dB & 10 dB  & 0 dB &  20 dB & 10 dB  & 0 dB&  20 dB & 10 dB  & 0 dB &  20 dB & 10 dB  & 0 dB\\
     \hline
     \hline
     \multirow{4}{*}{G.729B} & White   & 12.38  & 17.60 & 29.47 & 11.96 & 17.93 & 28.83 & 5.47  & 7.40 & 9.84 & 5.30  & 7.46 & 9.75\\
     \cline{2-14}
     &Pink     & 11.93  & 15.92 & 25.81 & 11.67 & 15.65 & 24.84 & 5.37  & 7.02 & 9.44 & 5.16  & 6.72 & 9.06\\
     \cline{2-14}
     &Volvo    & 12.50  & 13.07 & 15.22 & 12.37 & 12.40 & 13.44 & 5.54   & 6.01 & 6.82 & 5.41  & 5.39 & 5.97\\
     \cline{2-14}
     &Babble   & 12.07  & 16.66 & 26.86 & 11.26 & 15.15 & 25.18 & 5.63  & 7.34 & 9.55 & 5.01  & 6.73 & 9.36\\
     \cline{2-14}
     &Factory  & 12.03  & 16.09 & 25.01 & 11.43 & 14.92 & 23.63 & 5.45  & 7.06 & 9.40 & 5.20  & 6.65 & 8.82\\
     \hline
     \hline
     \multirow{4}{*}{SMVAD} & White   & 12.26  & 17.63 & 27.86 & 11.97 & 17.13  & 28.43 & 5.59  & 7.57  & 9.81 & 5.33  & 7.51 & 9.76\\
     \cline{2-14}
     &Pink     & 11.90  & 15.93 & 25.41 & 11.46 & 15.38 & 24.81 & 5.30  & 7.00 & 9.47 & 5.08  & 6.69  & 9.18\\
     \cline{2-14}
     &Volvo    & 12.48  & 13.21 & 15.05 & 12.54 & 12.30 & 13.44 & 5.53  & 5.93 & 6.75  &  5.36 & 5.39 & 6.01\\
     \cline{2-14}
     &Babble   & 11.87  & 17.10 & 28.29 & 11.27 & 15.82 & 25.95 &  5.65 & 7.51 & 9.65  & 5.12  & 6.79 & 9.58\\
     \cline{2-14}
     &Factory  & 11.83  & 15.66 & 24.04 & 11.47 & 14.95 & 22.83 & 5.34  & 6.93 &  9.15 & 5.11  & 6.55 & 8.54\\
     \hline
     \hline
     \multirow{4}{*}{HNPNR} & White   & 12.44  & 17.90 & 35.20  & 11.70 &  17.40 & 35.40 & 5.85  & 8.08 & 10.00  & 5.62  & 7.96 & 10.00\\
     \cline{2-14}
     &Pink     & 11.53  & 15.96 & 31.48 & 11.40 & 15.22 & 30.44 & 5.69  & 7.59 & 10.00 & 5.49  & 7.18 & 10.00\\
     \cline{2-14}
     &Volvo    & 10.86  & 11.79 & 14.15 & 11.06 & 11.36 & 12.64 & 5.24  & 5.92 & 7.03 & 5.17  & 5.40  & 6.20\\
     \cline{2-14}
     &Babble   & 11.20  & 15.69 & 26.28 & 10.56 & 14.21 & 25.27 & 5.61  & 7.38 & 10.00 &  5.31 & 6.74 & 10.00\\
     \cline{2-14}
     &Factory  & 11.23  & 15.45 & 28.74 & 10.86 & 14.82 & 27.96 & 5.56  & 7.45 & 10.00 & 5.36  & 7.01 & 10.00\\
     \hline
     \hline
     \multirow{4}{*}{MEBTS} & White   & 14.28  & 22.93 & 34.97 & 12.88 & 21.56 & 34.96 & 6.51   & 8.73 & 9.96 & 6.19 & 8.88 & 9.97\\
     \cline{2-14}
     &Pink     & 11.20  & 20.26 & 33.90 & 10.29 & 19.01 & 33.02 & 5.15  & 8.38 & 9.88 & 4.87 & 8.41 & 9.86\\
     \cline{2-14}
     &Volvo    & 10.12  & 11.19 & 12.71 & 9.82  & 10.06 & 11.20 & 4.82  & 5.13 & 6.01 & 4.49 & 4.69 & 5.24\\
     \cline{2-14}
     &Babble   & 10.60  & 17.20 & 32.41 & 9.69  & 15.46 & 31.48 & 5.07  & 7.59 & 9.82 & 4.61 & 7.12 & 9.82\\
     \cline{2-14}
     &Factory  & 11.00  & 17.67 & 32.12 & 9.86  & 17.36 & 31.14 & 5.07  & 7.97 & 9.78 & 4.70 & 7.84 & 9.78\\
     \hline
     \hline
     \multirow{4}{*}{AEBTS} & White   & 12.37  & 23.17 & 35.17 & 11.93 & 21.11 & 35.43 & 5.96  & 9.10 & 10.00 & 5.57 &  9.00 & 9.95\\
     \cline{2-14}
     &Pink     & 10.52  & 20.01 & 34.16  & 10.09 & 18.64 & 32.55  & 4.98  & 8.42 & 9.97 & 4.76 & 8.22 & 9.91\\
     \cline{2-14}
     &Volvo    & 10.16  & 10.67 & 12.41  & 9.76 & 10.22 & 11.22 & 4.73  & 5.09 & 5.78 & 4.60  &  4.76 & 5.16\\
     \cline{2-14}
     &Babble   & 10.56  & 15.52 & 32.35  & 9.86 & 14.21 & 31.41 & 4.88  & 7.07 & 9.91 & 4.59 & 6.59 & 9.82\\
     \cline{2-14}
     &Factory  & 10.60  & 16.93 & 32.58  & 9.76 & 16.45 & 30.51 & 4.90  & 7.70 & 9.88 & 4.64 & 7.29 & 9.77\\
     \hline
     \hline
     \multirow{4}{*}{UBGME} & White & 11.70  & 16.06  & 26.92 &  11.10  & 15.99 & 26.95  & 5.29  & 7.02 & 9.71 &  5.05  & 7.04 & 9.61 \\
     \cline{2-14}
     &Pink     & 10.93  & 14.45 & 23.97 & 10.50 & 13.95 & 23.24 & 5.08  & 6.52 & 9.29  & 4.85 & 6.29 & 9.10\\
     \cline{2-14}
     &Volvo    & 10.69  & 11.46 & 12.77 & 10.16 & 10.53 & 11.26 & 4.88  & 5.15 & 5.83 & 4.67 & 4.78 & 5.31\\
     \cline{2-14}
     &Babble   & 10.76  & 14.29 & 23.10 & 10.16 & 13.21 & 23.13 &  4.99 & 6.43 & 9.16 & 4.60 & 6.13 & 9.05\\
     \cline{2-14}
     &Factory  & 10.76  & 14.25 & 24.57 & 10.19 & 13.48 & 24.04 & 5.00  & 6.54 & 9.38 & 4.70 & 6.24 & 9.14\\
     \hline
     \hline
  \end{tabular}
  \label{TableNoiseSRE2002}
\end{table*}

\subsection{Results in Real-time Scenario}\label{Subsection:RealtimeCondition}
In Section~\ref{Subsection:MatchCondition} results are shown for clean condition whereas in Section~\ref{Subsection:MismatchCondition}, speaker recognition results are shown for noisy condition where in every case all the test speech segments are distorted with same noise of equal SNR. However, in real life, this kind of controlled environment is not replicated. In practice, most of the speech utterances are distorted with different type of noise of various SNR. In order to observe the performance of speaker recognition in this kind of situation, we have distorted different speech files of NIST SRE $2002$ with different noise of various SNR. The noise type is randomly chosen from the set of five noises used in the previous experiments. The SNR is randomly selected between $0$ to $40$. We call the distorted dataset as Distorted NIST SRE $2002$. As different speech files are collected for different environmental conditions, score normalization would be very effective here~\cite{Auckenthaler}. Here, we have applied $t$-normalization on raw log likelihood scores for generating final scores. The utterances for $t$ normalizations are chosen from training speech files of NIST SRE $2001$ i.e. $74$ male and $100$ female speech files are selected for $t$-normalization.

\begin{table*}
\renewcommand{\arraystretch}{1.3}
  \centering
  \caption{Speaker verification results in distorted NIST SRE 2002 for different speech activity detectors.}
  \begin{tabular}{|c||c|c|c|c||c|c|c|c|}
     \hline
     \hline
     \multirow{3}{*}{Method}   & \multicolumn{4}{|c|}{w/o $t$-norm} & \multicolumn{4}{|c|}{with $t$-norm}\\
     \cline{2-9}
                               & \multicolumn{2}{|c|}{EER (in \%)} & \multicolumn{2}{|c|}{minDCF $\times$ 100} & \multicolumn{2}{|c|}{EER (in \%)} & \multicolumn{2}{|c|}{minDCF $\times$ 100}\\
     \cline{2-9}
                               &  MFCC  & OBTC  &  MFCC  & OBTC&  MFCC  & OBTC  &  MFCC  & OBTC\\
     \hline
     \hline
     G.729B                    & 21.69 & 19.41 & 8.52 & 8.05 & 21.22 & 18.54 & 8.14 & 7.53\\
     \hline
     SMSAD                     & 22.09 & 19.51 & 8.65 & 8.16 & 21.35 & 18.74 & 8.37 & 7.72\\
     \hline
     HNPNR                     & 21.18 & 19.13 & 8.75 & 8.34 & 20.72 & 18.64 & 8.14 & 7.49\\
     \hline
     MEBTS                     & 23.73 & 22.79 & 9.94 & 9.93 & 23.57 & 22.26 & 9.96 & 9.97\\
     \hline
     AEBTS                     & 23.33 & 21.89 & 9.96 & 9.93 & 22.90 & 21.45 & 9.97 & 9.96\\
     \hline
     UBGME                     & 19.14 & 17.53 & 8.32 & 8.03 & 18.34 & 16.43 & 7.88 & 7.38\\
     \hline
     \hline
  \end{tabular}
  \label{TableReal}
\end{table*}

\section{Conclusion \& Future Work}\label{Section:Conclusion}
In this work, we briefly review some standard SAD techniques and their effects in speaker recognition performances. The performance of different techniques are evaluated on two NIST corpora: NIST SRE $2001$ and NIST SRE $2002$. SAD techniques like G.729B and SMSAD, which are very much accepted in speech coding and other applications, are shown to exhibit lower performance than even simple energy based speech activity detector. The experimental results show that speaker recognition system with two Gaussian modeling of log-energy based SAD is significantly better than other techniques for wide range of SNR. We have evaluated the performances using two cepstral features: standard MFCC and our previously proposed OBTC. It has been shown that OBTC based system is superior than MFCC for most of the cases.
\par
The performance of the UBGME based SAD appears to be suboptimal in the shown cases. However, this utterance wise bi Gaussian modeling of log-energy based approach can be further improved by investigating the followings subjects:
\begin{itemize}
  \item
  When a speech utterance gets distorted by noise, its different frequency bands are unequally affected (Fig.~\ref{FigFrequencyResponses}). Therefore, two Gaussian modeling of subband information could be used to improve the performance of the speech activity detector.

  \item
  The UBGME based approach uses only log-energy. However, bi Gaussian modeling of other parameters like entropy, spectral flatness parameter can be studied to extract robust speech frames.

  \item
  Decision fusion technique~\cite{ZaurNasibov} can be used to improve the performance further by combining multiple speech activity detector carrying supplementary information.
\end{itemize}

\bibliography{latexbib}

\end{document}